# Fast Desktop-Scale Extrusion Additive Manufacturing

**Jamison Go and A. John Hart***
Department of Mechanical Engineering and Laboratory for Manufacturing and Productivity, Massachusetts Institute of Technology, Cambridge, MA
*ajhart@mit.edu

## Abstract
Significant improvements to the production rate of additive manufacturing (AM) technologies are essential to their cost-effectiveness and competitiveness with traditional processing routes. Moreover, much faster AM processes, in combination with the geometric versatility of AM, will enable entirely new workflows for product design and customization.  We present the design and validation of a desktop-scale extrusion AM system that achieves far greater build rate than benchmarked commercial systems. This system, which we call 'FastFFF', is motivated by our recent analysis of the rate-limiting mechanisms to conventional fused filament fabrication (FFF) technology.  The FastFFF system mutually overcomes these limits, using a nut-feed extruder, laser-heated polymer liquefier, and servo-driven parallel gantry system to achieve high extrusion force, rapid filament heating, and fast gantry motion, respectively.  The extrusion and heating mechanisms are contained in a compact printhead that receives a threaded filament and augments conduction heat transfer with a fiber-coupled diode laser.  The prototype system achieves a volumetric build rate of ~127 cm$^3$/hr, which is ~7-fold greater than commercial desktop FFF systems, at comparable resolution; the maximum extrusion rate of the printhead is ~14-fold greater (282 cm$^3$/hr).  The performance limits of the printhead and motion systems are characterized, and the tradeoffs between build rate and resolution are assessed and discussed. The combination of high-speed motion and high deposition rate achieved by the FastFFF technology also poses challenges and opportunities for toolpath optimization and real-time deposition control.  High-speed desktop printing raises the possibility of new use cases and business models for AM—where handheld parts are built in minutes rather than hours. Adaptation of this system to high-temperature thermoplastics and filled composite resins, which require high extrusion forces, is also of interest.

# 1. Introduction

Extrusion-based additive manufacturing (AM), typically referred to as fused filament fabrication (FFF) or fused deposition modeling (FDM), is arguably the most versatile AM process due to its compatibility with a wide variety of thermoplastic materials and its wide range of operating length scales. The first commercial FDM systems were sold by Stratasys in 1991 [1], [2], and the rapid recent emergence of desktop AM technology is in large part due to the wide availability and declining cost of personal FFF/FDM systems. In FFF, a thermoplastic feedstock, which is typically stored as a spooled polymer filament, is fed to the print head where it is heated, melted, and forced through a nozzle. Molten polymer roads are laid in the horizontal plane to complete the cross-section of a part, and the stacking of consecutive cross-sections creates the final freestanding 3D object.

In addition to rapid prototyping, extrusion AM is increasingly used for creating functional parts and fixtures with ~0.1-1 millimeter-scale bead width ('small-bead' extrusion). Larger format printers such as the Big Area Additive Manufacturing (BAAM) system from Oak Ridge National Laboratory create objects on the meter scale, by extrusion of pelletized feedstock into beads with 0.1-1 centimeter width ('large-bead' extrusion) [3]. Objects created by large-bead extrusion AM include office furniture, vehicle bodies, and large tooling [4]. The adaptation of robotics to extrusion AM also increases the size and complexity of parts that can be created, for example enabling non-planar layers to be printed, and combining additive and subtractive operations in a single station or workcell [5]–[7].

Across the spectrum of extrusion AM, lower-cost thermoplastics such as ABS, PLA, and Nylon are commonly employed. Higher temperature thermoplastics such as ULTEM, polycarbonate, and PPSF are more frequently compatible with industrial systems, and are of growing interest for applications that demand higher strength and thermal stability such as tooling and aircraft ductwork. Composite blends, such as ABS with chopped or glass fiber, can offer much improved mechanical properties [8], yet present challenges due to their higher effective viscosity and abrasiveness.

For all AM technologies, improvements in production rate, and optimization of intrinsic tradeoffs among quality, cost, and rate, are essential to accelerate adoption and industrialization. Higher rate AM machines are attractive to both service bureaus and direct users (e.g., engineers/consumers). Reduced build times allow for higher production rates which means greater returns on capital purchases and lower production costs per part, favoring the economic viability of AM as a short-run and custom manufacturing technique. For direct users, faster prototyping of parts reduces time spent on concept development, accelerating the design-build-test loop. Further, desktop AM with significantly higher build speed (~10X or greater) may open up entirely new opportunities for AM. For instance, if small parts could be fabricated by AM in ~minutes rather than ~hours, specialized AM machines could become commonplace in urgent medical practice, could be used to produce spare parts on-demand in repair shops, or could simply be used to realize a physical prototype during the duration of a coffee break.

Recently, we found that that the volumetric build rate of small-bead extrusion AM can be limited by the maximum force applied by the material feed mechanism, the rate of heat transfer from the liquefier wall into the filament core, and/or the performance of series gantries controlled by open-loop stepper motors [9]. Therefore, the maximum build rate—which is comparable for both consumer desktop (~$500) and professional (~$100,000) FFF systems—can be related to the design and performance of the machine elements, as coupled to the requirements for heating and flow of the thermoplastic through the extrusion printhead. In Fig. 1, we summarize the trade-space between volumetric build rate and resolution (defined as the extrusion nozzle diameter) predicted by this analysis, and validated by direct measurements of the build rates of several commercial FFF machines. Each of the three limits can bound the achievable rate-resolution performance of an FFF system, and therefore, machine designs that simultaneously achieve high extrusion force, rapid heat transfer to the polymer, and rapid yet accurate positioning of the printhead, are needed to significantly increase the build rate of FFF.

Within the bounds of Fig. 1, substantial improvements in these performance metrics could enable at least a 5-10X increase in the FFF build rate without sacrificing part detail (resolution). Here, we present a fast extrusion AM technology which we refer to as FastFFF. The prototype machine described in this work mutually overcomes the module-level performance limits, enabling volumetric build rates at least an order of magnitude greater than the benchmarked commercial desktop systems.

## 2. FastFFF System

Fast extrusion AM is achieved by integrating a novel high-throughput printhead, described below, with a high-speed gantry system driven by DC servo motors. The desktop FastFFF system, with peripheral electronics, is shown in Fig. 2a. A threaded filament is provided to the printhead, where it is fed forward using a rotating nut, which we call the 'nut-feed' mechanism. Downstream of the nut-feed mechanism, the filament passes through a small quartz chamber coupled to a diode laser, surrounded by a gold reflector. The filament is heated to an intermediate temperature by the laser, and then passes into a heated tapered liquefier where it finally reaches the melt temperature and exits the print nozzle. As shown below, the combination of high extrusion force and rapid heat transfer is critical to achieve significantly greater volumetric build rate. For example, a small twisted pentagonal prism cup is printed in 376 seconds, as shown in Fig. 2b.

*2.1 Nut-feed filament drive*

In commercial desktop and professional FFF systems, the filament is advanced using a pinchwheel mechanism, whose force capability is limited by the depth of grooves imprinted into the smooth filament by the textured drive wheel, which in turn determines the area of traction. When the required extrusion force exceeds the shear strength of the polymer over the traction area, the filament fails at the pinchwheel interface, and traction is lost [9]. As a result, FFF extrusion mechanisms operate at a fraction of their maximum force capacity, and the low transmission ratio between the rotation rate of the pinchwheel mechanism and the exit velocity of the filament (due to the significant area reduction of the nozzle) limits the fidelity and performance of extrusion rate.

In the FastFFF system, this limit is overcome using a rotating nut to drive a threaded filament (Fig. 2c), analogous to a leadscrew linear actuator. Therefore, the filament is pushed forward linearly by the interaction of matching threads on the outside of the filament and inside of the drive nut. The drive nut rotates axially relative to the filament. The feed rate ($\dot{z}$) is related to the thread pitch ($p$) and the angular velocity of the feed nut ($\dot{\theta}$) as

$$\dot{z} = \frac{p}{2\pi}\dot{\theta} \quad (1)$$

The drive nut is constrained between bearing elements such that only axial rotation is possible; the filament is similarly constrained by a pair of rollers (with rotation axis perpendicular to the filament axis) to allow only linear translation. The nut-feed mechanism also introduces a natural 'gearbox' enabling fine control over the feed rate by the motor rotation rate, and the threads, which wrap around the entire filament circumference, enable a much greater axial force to be generated as the filament is driven into the heater.

The ABS filament is cut with a standard 4-40 thread (Craftech Ind.), which has a triangular profile (Fig. S1, Table S1) [10]. The maximum linear force capability can be estimated according to the root area of the filament that is in shear ($A_s$) during axial translation of the filament through the feed nut, multiplied by the approximate shear strength for ductile materials [10], [11]:

$$A_s \cong \frac{\pi l_e d_{max}}{p}\left[\frac{p}{2} + \frac{1}{\sqrt{3}}(d_{max} - d_p)\right] \quad (2)$$

$$F_{max} \cong \sqrt{3}\sigma_{uts}\left(\frac{\pi l_e d_{max}}{p}\left[\frac{p}{2} + \frac{1}{\sqrt{3}}(d_{max} - d_p)\right]\right) \quad (3)$$

Thus, the filament major diameter ($d_{max}$), filament pitch diameter ($d_p$), length of engagement with the drive nut ($l_e$), pitch, and approximate shear strength ($\sigma_s \cong \sqrt{3}\sigma_{uts}$) determine the force capacity. A larger diameter, deeper thread, finer pitch, longer engagement length, and greater material strength result in greater axial force. However, many of these same attributes inversely affect required motor torque by introducing greater friction between the filament and nut. Increased filament diameter and coarser pitch for example will increase friction which demands higher torque from the extrusion motor and places torsion on the filament body. Therefore, optimization of the thread profile and/or coating of the nut threads to reduce friction would enable higher transmission efficiency and lower stresses in the filament.

The tensile force capacity of the feed mechanism was tested using a sample length of filament held in a threaded nut (Fig. S2); for a single thread engagement, shear failure at the root occurred at 109 N axial force (averaged over strain rates of 0.1-10 mm/s), which compares well to the estimate of 95 N calculated using the thread geometry and the tensile strength of ABS ($\sigma_{uts}$ = 37 MPa). During operation of the extruder at the highest forces (>150 N), we find that the filament fails in torsion, due to the high friction mentioned above. Also, the threads experience combined normal and shear loading during operation, and the mechanics of the plastic will depend on the strain rate [12], making it difficult to precisely estimate the maximum extrusion force capability. Nevertheless, the measured axial shear strength of a single 4-40 thread ($F$ = 109 N; $l_e$ = 0.64 mm) far exceeds the maximum force measured for the traditional pinch wheel extruder mechanism (59 N) [9].

*2.2 Laser-assisted heater*

To achieve rapid extrusion AM it is also necessary to heat to the polymer to the target temperature much more rapidly than in conventional designs that rely on thermal conduction between the moving filament and a heated block (the 'liquefier'). In conventional designs, the filament first moves through a channel matching its nominal diameter, and ideally, its core approaches the melt temperature prior to entering the nozzle where it is reduced significantly in diameter. Incomplete heating of the polymer that exits the nozzle adversely influences adhesion of the printed material to the previous layer. The low thermal conductivity of the polymer also results in significant radial thermal gradients, and the required extrusion force increases sharply with feed rate if the filament core is insufficiently heated when reaching the nozzle reduction. Therefore, the coupling between heat transfer and flow in the liquefier determines the required heater length, which in turn influences the required force. An ideal design would allow the filament to be heated uniformly and rapidly through its cross-section, enabling smooth extrusion at high volumetric feed rates.

In our approach, the moving filament is first heated to an intermediate temperature by a near-infrared diode laser ($\lambda$ = 808 nm, DILAS-MINI, 50 W maximum power), then travels into liquefier where it is heated further to the melt temperature and then extruded through a nozzle (Fig. 2d). Infrared heating of polymers is widely used in manufacturing [13], wherein the spectral properties of the polymer and heat source determine the penetration depth of thermal radiation and the efficiency of coupling. Assuming that the laser fully penetrates the filament, heating is volumetric rather than surface-based, and the non-contact heat transfer is both rapid and frictionless. To achieve this, the laser wavelength must be chosen based on the optical properties of the filament and any additives such as coloring pigments.

The laser is coupled to the printhead via a sheathed optical fiber, and the filament is exposed to the laser within a small cylindrical chamber. The collimated laser beam (~3 mm diameter) enters the chamber at an angle of 5° from the perpendicular plane; this orientation is chosen to establish multiple reflections of the laser within the chamber, giving a radially uniform intensity distribution and encouraging uniform volumetric heating as determined by the absorption spectrum of the ABS (Fig. S3). The filament passes through a custom-machined optical-grade quartz cell (8 mm OD, 3 mm ID, 10 mm long) which features 90° chamfers on each end to co-linearly mate and seal between adjacent IR chamber structures. Aside from the laser beam entry port, the quartz cell is surrounded by gold foil which ensures reflectivity of the IR energy with minimal losses. These elements are fully enclosed by an aluminum housing for safety and rigidity.

The filament temperature upon exit from the laser heating chamber will therefore be related to the incident power and the filament feed rate; this was characterized by thermal imaging (FLIR T430sc) of the filament upon output from the chamber (Fig. 3a). To avoid overheating of the filament surface, the laser power is programmed to be proportional to the filament feed rate (Fig. S4), resulting in the output curves of average surface temperature versus feed rate shown in Fig. 3b. The values correspond to the peak temperature of the filament at the exit of the IR chamber as observed by the camera, and below this point the filament temperature drops gradually as it is fed into open air.

Immediately downstream of the laser, the filament enters a modified conduction heater, whose role is to bring the pre-heated filament to the melt temperature, and then constrict the molten polymer through the nozzle. A tapered channel geometry is used to first deform the textured (threaded) surface of the pre-heated filament to enable good thermal contact with the liquefier walls which are maintained at the target melt temperature ($T_m$ = 260 °C). The channel is machined using a small tapered end mill; for the 4-40 filament, the channel tapers from 2.8 mm to 2.0 mm diameter along a distance of 11.5 mm.

The performance of the printhead in enabling high rate extrusion of molten thermoplastic is validated by measurements of the force required versus filament feed rate (Fig. 3c). A load cell was placed between the extruder and heating (laser + conduction) sections of the print head, as shown in Fig. S5. Without the IR laser, the printhead was able to extrude material at rates of 141 and 177 cm$^3$/hr for nozzle diameters of 0.5 mm and 1.0 mm respectively at which the measured force approached ~160 N, and the filament subsequently failed in torsion. The addition of the IR laser shifts the curves right with increasing laser power, allowing higher volumetric rates to be achieved at the same force. For instance, with the laser power set at an equivalent rating of 2.5 J/mm (power divided by filament feed rate), a maximum extrusion rate of 282 cm$^3$/hr is reached for a 1.0 mm nozzle. Under otherwise identical conditions, the force required for the 0.5 mm nozzle is always greater than the 1.0 mm nozzle, due to the greater viscous friction from the smaller nozzle diameter. The span of Fig. 3 shows that the rate-resolution space of the Fast-FFF printhead far surpasses the previously measured limits of commercial desktop FFF systems. For example the Stratasys Mojo operates at a volumetric infill rate of 21 cm$^3$/hr, corresponding to an estimated extrusion force of 30 N [9].

Due to the complexity of the chamber and filament geometry, a complete predictive model of the pre-heater is beyond scope of this paper; however, first-order calculations show that the efficiency of coupling between the laser and filament has not achieved the highest expected efficiency. For example, assuming the measured surface temperature represents the mean temperature of the filament, approximately 50% of the laser power is transferred to the filament at the highest volumetric rates. We also find a sub-linear relationship between laser power and filament temperature at constant feed rate, suggesting that, in spite of the enclosed chamber design, radiative heat loss to the surroundings may be significant, especially between the exit of the laser chamber and the entry of the tapered channel. Further investigation of these points would enable even greater extrusion throughput.

*2.3 High-speed gantry system*

In order to take advantage of the high extrusion rate of the printhead, a positioning mechanism with high speed and high dynamic accuracy is also required. An H-frame architecture was chosen (Fig. 1a, Fig. S6), because it enables the motors to be mounted to the fixed machine frame, reducing the moving mass while allowing higher torque motors to be used therefore giving further improved dynamic performance. The H-frame is powered by a pair of closed-loop controlled servomotors (Estun EMJ-04 Servomotor; Estun PRO-E-04B Servodrive). The motors are connected to the motion stage (which holds the printhead) through a single, continuous urethane-kevlar miniature extra-light (MXL) timing belt (Gates 962MXLL025). Rigid aluminum bearing blocks and hardened steel rods are used to provide high stiffness and smooth motion of the stage.

The kinematics and dynamic performance of the H-frame architecture are well-established in literature [14]. The position of the output stage (x, y) is related to the angular position of each motor ($\varphi_1$, $\varphi_2$) according to Eq. 4.

$$\Delta x = -\frac{1}{2}r\Delta\varphi_1 + \frac{1}{2}r\varphi_2; \quad \Delta y = -\frac{1}{2}r\Delta\varphi_1 - \frac{1}{2}r\varphi_2 \quad (4)$$

Because both motors must be driven to cause gantry motion along either orthogonal machine axis, higher accelerations can be achieved for motions along the axis directions. However, the intrinsic anisotropic stiffness of the belt path, and the coupled kinematics of the H-frame, must be considered in evaluating the static and dynamic positioning accuracy.

To begin, we may approximate the required gantry speed ($V_{print}$) in terms of the volumetric build (extrusion) rate (BR), nozzle diameter ($d_n$) and layer thickness ($h$), $V_{print} \cong BR/(hd_n)$. The build rate is simply proportional to the filament feed rate ($V_f$), $BR = \pi V_f d_f^2 / 4$, where $d_f$ is the mean diameter of the filament ($d_f$ = 2.4 mm for 4-40 thread). For example, corresponding to the maximum filament feed rates in Fig. 3c, $V_{print}$ = 480 mm/s at ($d_n$, $h$, $V_f$) = (0.5 mm, 0.2 mm, 10 mm/s); and $V_{print}$ = 385 mm/s at ($d_n$, $h$, $V_f$) = (1.0 mm, 0.2 mm, 16 mm/s). In addition, the gantry must be able to position the printhead with repeatability that, as a rule of thumb, is ~5-10X smaller than the lateral printing resolution. The selected nozzle diameters of 0.5 and 1.0 mm, which approximate the lateral resolution of the print, impose a specification of ~0.05-0.10 mm repeatability on the motion system. In addition, high accelerations are required to reach the high steady-state build rates, therefore introducing significant forces on the gantry and transmission elements.

The static and dynamic accuracy and repeatability of the H-frame gantry was measured along each axis ($x, y$) using a laser displacement sensor (Keyence LKG-152). Static accuracy was measured by moving the gantry (at a constant velocity command $V_{print}$ = 280 mm/s) by a specified distance, and comparing its final position to the command position after coming to rest. Dynamic accuracy was measured, versus speed, by the endpoint positions the printhead when commanded to execute a continuous back-and-forth motion. The repeatability is defined as the standard deviation of the respected errors, sampled over a series of identical measurements.

To examine dynamic positioning, the gantry was commanded to move repetitively between two positions along the axis at gantry speeds ranging from 35 to 280 mm/s at a prescribed acceleration of 10,000 mm/s$^2$. An example dataset shows the X and Y position of the gantry during this sequence, commanded at a rate of 245 mm/s (Fig. S7a). The error in gantry position was recorded and plotted versus gantry speed (Fig. S7b). The Y axis error does not show a consistent relationship with speed, and ranges between 0.05-0.15 mm over the tested speed range. However, the dynamic X axis error increases with gantry speed, starting at -0.17 mm at 35 mm/s and increasing to a maximum 0.32 mm at 210 mm/s. The static positional accuracy (Fig. S7c, d) was determined to fall within a band of ±0.1 mm.

The anisotropy of accuracy, which primarily influences dynamic positioning, is rationalized based on the differences in stiffness and mechanical complexity between the X and Y axis. The Y axis, upon which the bridge moves, includes shorter sections of belting. For Y motion, each servo motor applies an equal tension to the bridge. Conversely, X motion requires translation of the entire length of the belt, along with components both on the frame and bridge, and is susceptible to racking forces from belt tensions in opposite directions. The dependence of X position accuracy on speed is explained by friction losses from the additional mechanical elements which must move on the bridge and the lower axis stiffness excited by the racking forces. At low speeds, friction losses cause the print head position to fall short of the commanded position; at higher speeds, momentum from the printhead and stiffness dominate which causes position overshoot on high changes in acceleration. In addition, positioning accuracy can be influenced by integer rounding errors in the firmware regarding gantry kinematics or commanding positions between the resolutions of the actuators.

Importantly, the servomotors used in the H-frame gantry provide significantly more performance than the stepper motors used in typical FFF systems. For instance, the chosen servomotor (EMJ-04) has a stall torque of 3.8 N-m, which is ~10-fold greater than the stall torque measured for stepper motors used in a typical desktop FFF system, and in part due to closed loop control, can operate at a much higher rotation rate [9]. There is a significant difference in mass (2.50 kg compared to 0.15 kg) between the servomotors and stepper motors, which is overcome using the H-frame design to mount the motors to the stationary frame. In the current configuration, electronic gearing was implemented and a pulley system was used, resulting in a motor-gantry transmission ratio of 81.3 mm/rev, and a maximum calculated linear speed of approximately 0.9 m/s. We have safely commanded the gantry to ~0.7 m/s before excessive vibration and loss of positioning accuracy. Therefore, comparing to the printhead speeds required to meet the measured maximum extrusion rates (Fig. 3), the gantry motion system will not be the rate limit to the FastFFF system.

## 3. High-Speed Printing

The module-level performance of the FastFFF system translates into the capability to build thermoplastic objects significantly faster than all known systems that operate on a comparable resolution scale. In Fig. 4a we show a variety of test objects printed using ABS which include a miniature chair, a model of MIT's Building 10, eyeglass frames, a helical bevel gear (Video: http://bit.ly/2fAMPCM) with an internal taper interface, and a small spiral cup. These objects were printed using either a 0.5 mm or 1.0 mm nozzle with a layer height of 0.2 mm (Fig. 4b), using settings chosen to compromise between build rate and overall quality (Table S2). Table S3 provides a summary of statistics of these parts including their volumes and average volumetric print rate.

To realize high-speed printing with the prototype system, the maximum gantry speed was calculated to be 0.9 m/s, according to the maximum pulse frequency of the open-source 3D printer motherboard (RAMBo), the dimensions of the gantry hardware, and electronic gearing available from the servo drives. These parameters were input into the open-source slicing engine (CURA) and adjusted based on the empirical results of test prints. For example, we found it necessary to use a slower speed (up to 100 mm/s) on the first layer of each part to establish good adhesion with the build surface. The external and internal perimeters are printed at 200 or 250 mm/s, respectively, to give improved surface quality especially at corners. Although the infill and traverse speeds were specified at the calculated maximum, the slicing engine will determine the actual infill rate. For example, a portion of the part infill may be printed slower than 600 mm/s on 'fast' settings if the layer has a total of less than one second to cool; other limitations include maximum allowable acceleration rates and maximum filament feed rates. As a result of these different local extrusion rates, there is a non-linear correlation between part volume and print time. Therefore, parts with a large surface area, such as thin shells, will have an average build rate representative of the perimeter settings. On the other hand, larger parts and/or simple geometries with a significant fraction of infill will have greater average build rate, which will approach the volumetric infill rate.

The accuracy of material placement importantly influences the fidelity of features and the mechanical properties that arise from the part microstructure. The layer thickness is found to be highly uniform, as shown in a representative image of a part sidewall (Fig. 4b). To examine in-plane geometry effects, we constructed a simple test part comprising a stepped geometry with varying convex and concave radii of curvature (Fig. 4c, d). Comparing this part printed under slow (infill = 150 mm/s, perimeter = 150 mm/s) and fast speed settings (Table S2), we note that the straightness of the edges and radius of curvature of the corners were nearly identical between two parts, except for the sharp corners (fillet radius = 0 mm) in which the fast part experienced overshoot on the external corner (the perimeter swings too wide after a corner, rounded edge) and overfill in the internal corner (extra material beads up on the external surface). However, the infill of the fast part shows greater irregularity (seen more clearly in Fig. S8), which is exacerbated by the short lateral dimensions of the infill of the test part, in combination with the high accelerations of the fast print settings.

The tensile mechanical properties of FastFFF parts were examined using dogbone specimens (Fig. 4e), tested until breakage according to ASTM standard D638 [15]. These specimens were printed using 100% nominal infill, yet the slicing software limited us to creating specimens with alternating (45°/135°) layer orientations. As a result, while the measured tensile strength of ~32 MPa and ultimate strain of 8.8% is acceptable, the modulus of the parts (~0.5 GPa) is 2-5 fold lower than expected for ABS in tension [16]. While there is need for further understanding and optimization, the consistency of these values over several samples verifies the robust structural mechanics of ABS parts printed at high infill speeds.

In conventional FFF, the heat transfers between the hot material being deposited and the underlying layer influences the local part microstructure, interlayer adhesion, and overall dimensional quality due to thermal stresses and deformation. As a result, it is desirable for the underlying layer, and the part as a whole, to remain at a constant temperature regardless of the trajectory history of the print. In the present case of fast printing, the rate of heating delivered to the part via the deposited material is greater than standard FFF, and the layer print duration is much shorter. As a result, we found it critical to locally cool the part surface in the vicinity of printing, to avoid excessive heat accumulation that led to distortion. This was achieved using a cage-type cooling fan impinging on the newly deposited material from the right side of the nozzle. However, we found that the local part quality varied according to the local direction of airflow relative to the direction of build, indicating that thermal conditions were not optimized. Therefore, to achieve high quality fast printing, especially for parts with complex cross-sections creating widely different thermal boundary conditions and local infill trajectories, omnidirectional air cooling is likely needed. Within a heated build chamber, this could be achieved by locally directing the ambient air over the surface of the part, while allowing a strong interface to form between adjacent layers. This is a compromise between the intrinsic high-speed deposition which demands a higher cooling rate for dimensional control of small parts, and the intrinsic time needed for the polymer to spread at the layer interface.

In Fig. 5, we compare the operating performance of the FastFFF system, in terms of build rate and resolution, to the previously benchmarked suite of desktop and commercial FFF systems [9]. To account for the different nozzle diameters and layer heights of each machine, a resolution ($R$) metric is defined, $R = \sqrt{d_n h}$. Using each machine, we printed the same part—a triangular prism (74.5 mm side lengths, 20 mm tall)—and measured the time-averaged build rate [9]. The FastFFF system achieves 85 cm$^3$/hr with ($d_n$, $h$) = (0.5, 0.2) and 127 cm$^3$/hr with ($d_n$, $h$) = (1.0, 0.2). This is 4.7 and 7 times faster, respectively than the build rate of the Strasys Mojo (18 cm$^3$/hr), up to 10 times faster than the Zortrax M200. The FastFFF system also compares well against the Stratasys Fortus 360mc (not shown), an industrial FFF system which averaged a build rate of 60.9 cm$^3$/hr and has a significantly larger printhead (albeit with a conventional design) and motion system. Moreover, the maximum volumetric rate determined by the printhead test is 282 cm$^3$/hr, which is 14-fold greater than the infill rate directly measured for the Stratasys Mojo; therefore, it is likely that further optimization of the toolpath and nut-feed design can enable the average build rate to approach or even exceed this value.

To establish commercial viability of this technology, we must further understand these and other issues that are introduced by the high-speed deposition conditions.  First and foremost, the microstructure of the part must be accurately controlled, both within the part interior and at critical geometric features at corners.  The present results demonstrate the ability of our process to achieve very high build rates, and good overall form and surface quality; however, it is clear that the extrusion and motion dynamics have not been optimized.  In traditional FFF, infill defects are common near corner radii because it is difficult to control the material flow quickly and thus the deposition rate does not properly match the acceleration profile of the gantry; overlapping road deposition patterns can fill internal voids but can simultaneously create areas of overfill in which parts lose dimensional accuracy [17]. Generation of acute features or infill of corners smaller than the road width still present a challenge because they cannot be created without overfilling [18], [19].  At high build rates, it may be necessary to couple the extrusion rate to the local motion speed; in other words, the machine would need to slow the extrusion rate when approaching a corner, to avoid overfilling.  Doing so is challenging due to the intrinsic flow dynamics of the polymer, which impose a time delay between the mechanical action of the extruder and the flow at the nozzle [17], [20].  In fact, an earlier configuration of our system used a Bowden-feed configuration where the nut-feed extruder was mounted to the motion frame; this was not pursued because the force in the tube-fed filament prevented rapid modulation of the polymer flow at the nozzle and led to significant excess deposition at corners and toolpath path endpoints.  This issue is further exacerbated by die swelling, which is the expansion of the material upon exit from the nozzle due to release of stored mechanical energy; studies have shown this can result in up to a 30% increase in the bead diameter, depending on the shear rate [17], and is certainly present in our system. Die swelling in particular often results in extra deposited material on layer transitions or between non-contiguous bodies in a cross section of a part; a better technique than simply retracting filament to relieve liquefier pressure is needed for precise flow control in FFF printers.

Last, it is well-known that the cross-sectional shape of the deposited material can be varied by modulating the velocity of the print head in conjunction with the material deposition rate [19], [21].  This, in combination with the capability for high volumetric build rates, may enable new slicing and toolpath strategies to achieve fast builds along with locally tailored resolution.  For example, it would be interesting to investigate adaptive slicing strategies that vary the thickness of each layer according to the local curvature, or the use of larger bead dimensions in the part interior while maintaining fine bead dimensions for surface details [22], [23].

## 5. Conclusions

In summary, we have demonstrated a machine for high-speed extrusion AM of polymers, achieving ~10-fold greater build rate than commercial desktop FFF systems.  This system serves an example of how AM technologies may be improved by understanding and overcoming performance-limiting aspects.  To achieve the true potential of this approach, it will be necessary to employ improved toolpath planning algorithms, and further evaluate the coupling between high-speed motion dynamics and extrusion dynamics.  Nevertheless, the capability to fabricate handheld objects in 5-10 minutes allows us to conceptualize uses for immediate prototyping, and for scaling of the technology to larger dimensions and/or to other materials such as high-temperature thermoplastics and thermoplastic composites.

# 6. Acknowledgements

This research was supported by a grant from Lockheed Martin Corporation, managed by Dr. Padraig Moloney. We thank the MIT International Design Centre (IDC) and MIT MakerWorks for providing spaces for experimentation and fabrication. We also thank Prof. Martin Culpepper for loan of the infrared camera; Dr. Scott Schiffres and Amelia Helmick for assistance with measurements and materials characterization; and Stuart Baker, Adam Stevens, and Greg Dreifus for valuable discussions.

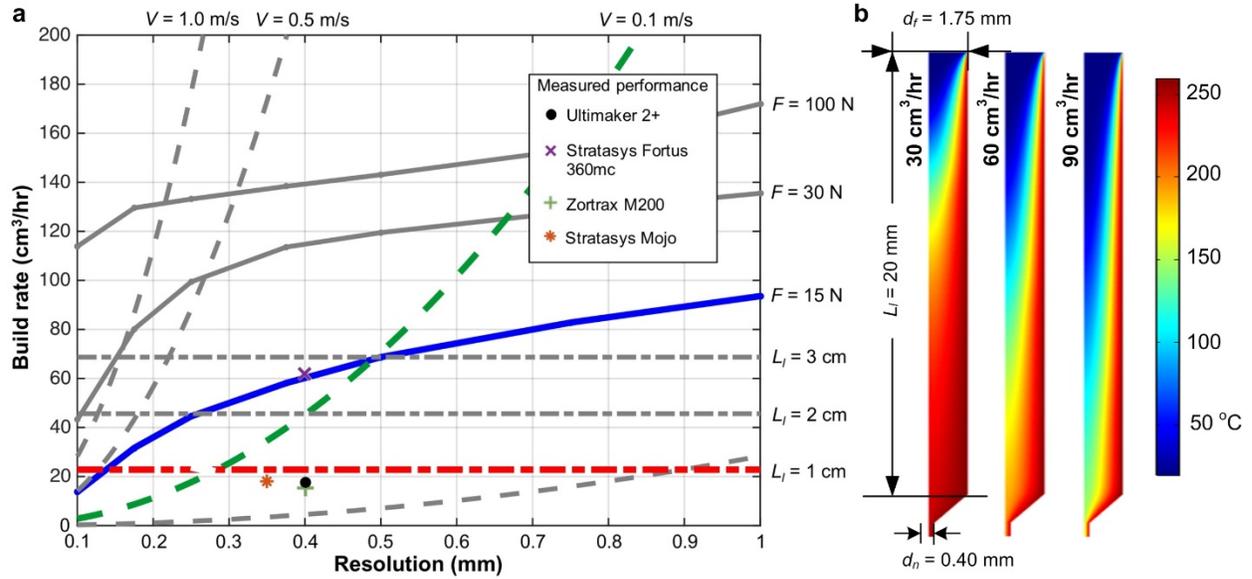

**Figure 1.** Summary of conventional FFF machine performance: (a) Performance space of FFF system build rate versus resolution as constrained by maximum gantry speed, extrusion force (corresponding to $L_l = 2$ cm), and liquefier length (with $T_{core,exit} = 0.9 T_{melt}$). The common area under the three curves defines the accessible performance for the system; selected desktop FFF systems are noted within the regime map; (b) temperature distribution within a conduction liquefier with volumetric rates of 30, 60, and 90 cm³/hr, demonstrating limited thermal penetration at higher feed rates.

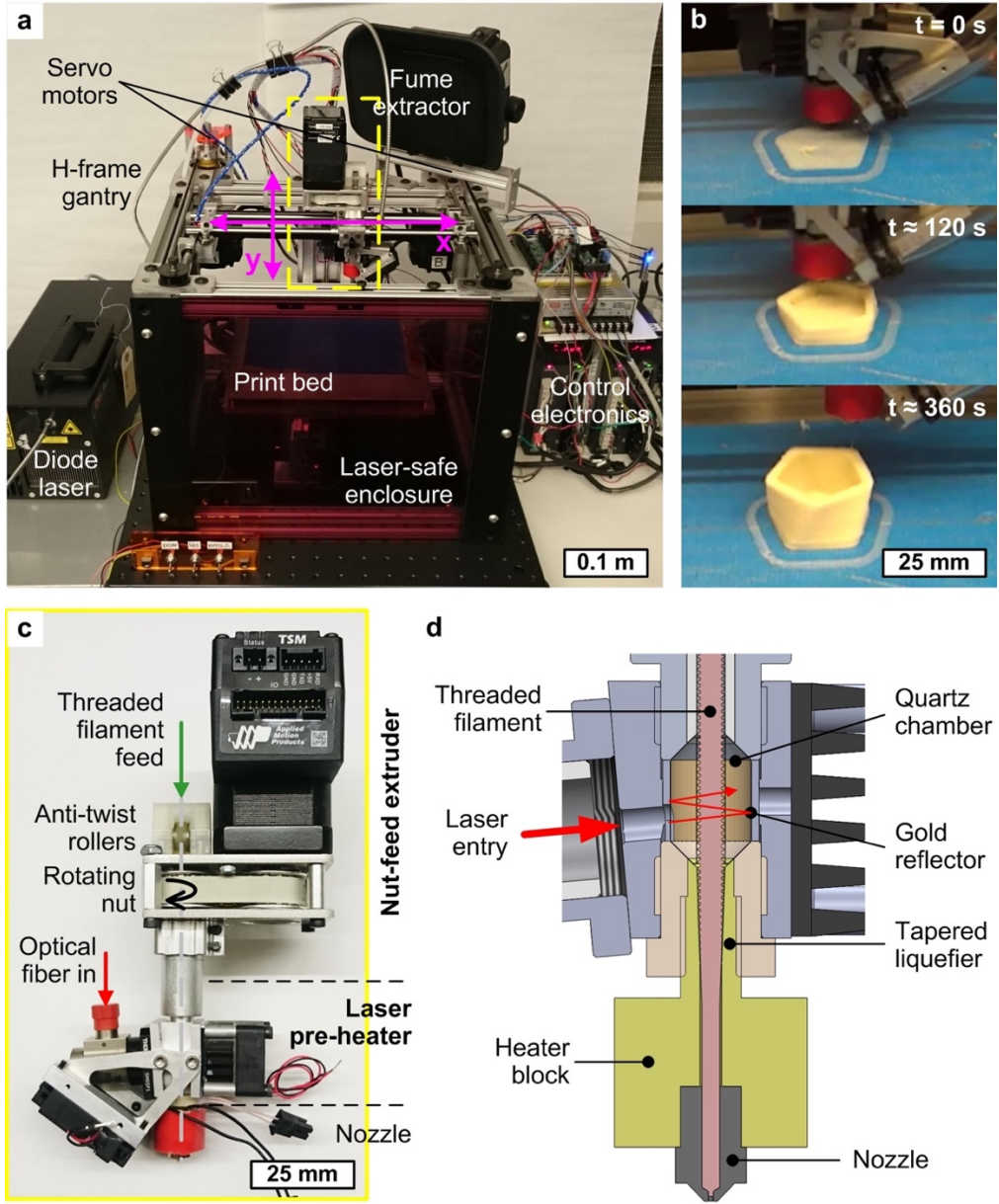

**Figure 2.** FastFFF system and high-throughput printhead design: (a) desktop FastFFF system (build volume 185×125×200 mm), showing H-frame gantry, diode laser with fiber coupling to printhead, and control electronics; (b) photographs during printing of a spiral cup using ABS feedstock; (c) photograph of printhead, showing nut-feed drive mechanism and laser-assisted hot end. (d) cutaway solid model of hot end, showing how laser is coupled to filament as it travels through a quartz chamber surrounded by a reflective gold foil. After the filament exits the laser heating chamber, it enters the conduction heating chamber, which is tapered to facilitate good thermal contact with the threaded surface of the filament.

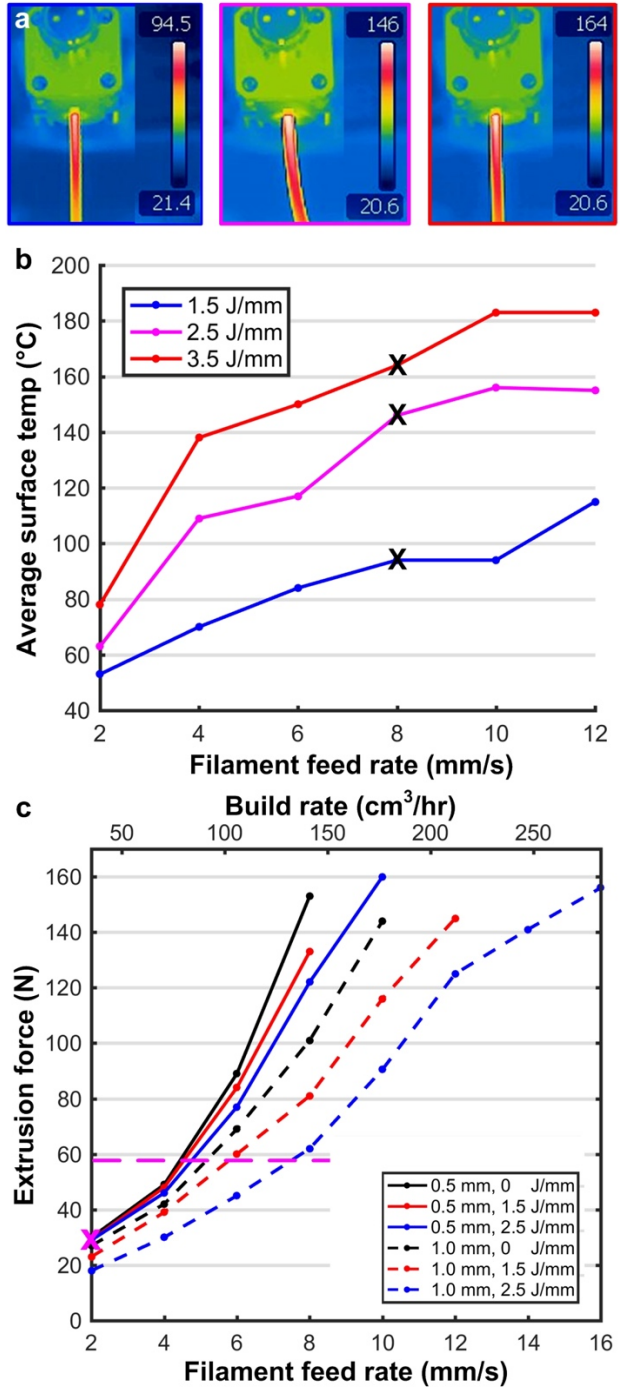

**Figure 3.** Evaluation of the IR-assisted heater and its influence on the force-rate relationship and maximum filament feed rate: (a) IR camera captures of filament exiting the IR heating chamber at 8 mm/s with power constants or 1.5, 2.5, 3.5 J/mm; (b) plot of the average surface temperature of filament exiting the IR heating chamber with increasing filament feed rate and power constants of 1.5, 2.5, and 3.5 J/mm; (c) steady-state extrusion force of the FastFFF printhead with increasing filament feed rate for nozzle diameters of 0.5 and 1.0 mm and power constants 0, 1.5, and 2.5 J/mm. The data shows that additional IR heating both reduces the force required to achieve a particular feed rate, and increases the maximum achievable feed rate before mechanical failure of the filament.

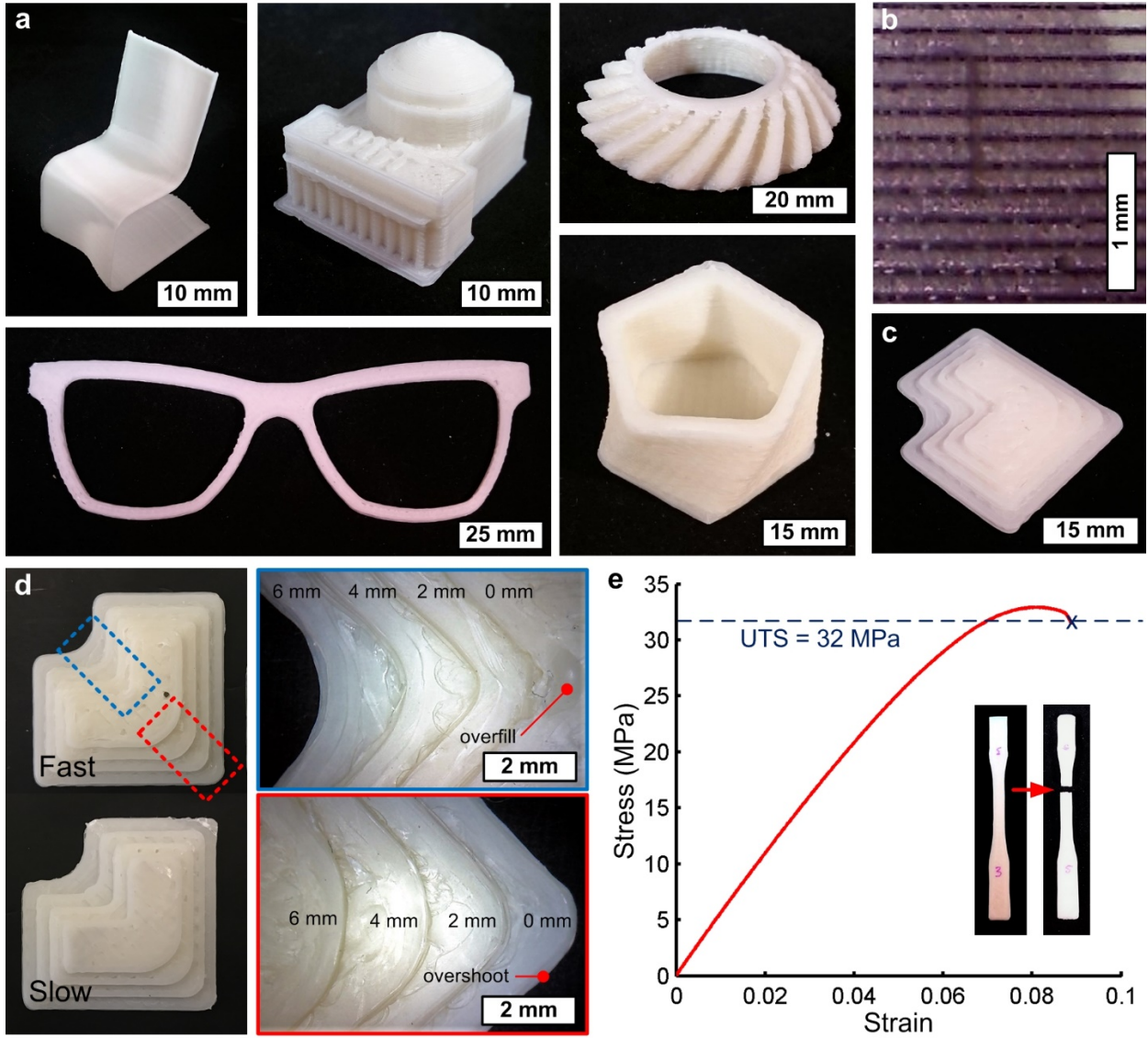

**Figure 4.** Collection of objects and test samples printed on the FastFFF printer: (a) a miniature chair, simplified model of building 10 at MIT, eyeglass frames, a spiral cup, and a helical bevel gear with internal taper printed on the FastFFF machine; (b) microscope image which verifies the layer height of 0.2 mm; (c) test part to evaluate corner fidelity; (d) comparison of fast and slow corner test parts demonstrates the ability to generate sharp corners at high speeds; (e) tensile test data for representative dogbone sample.

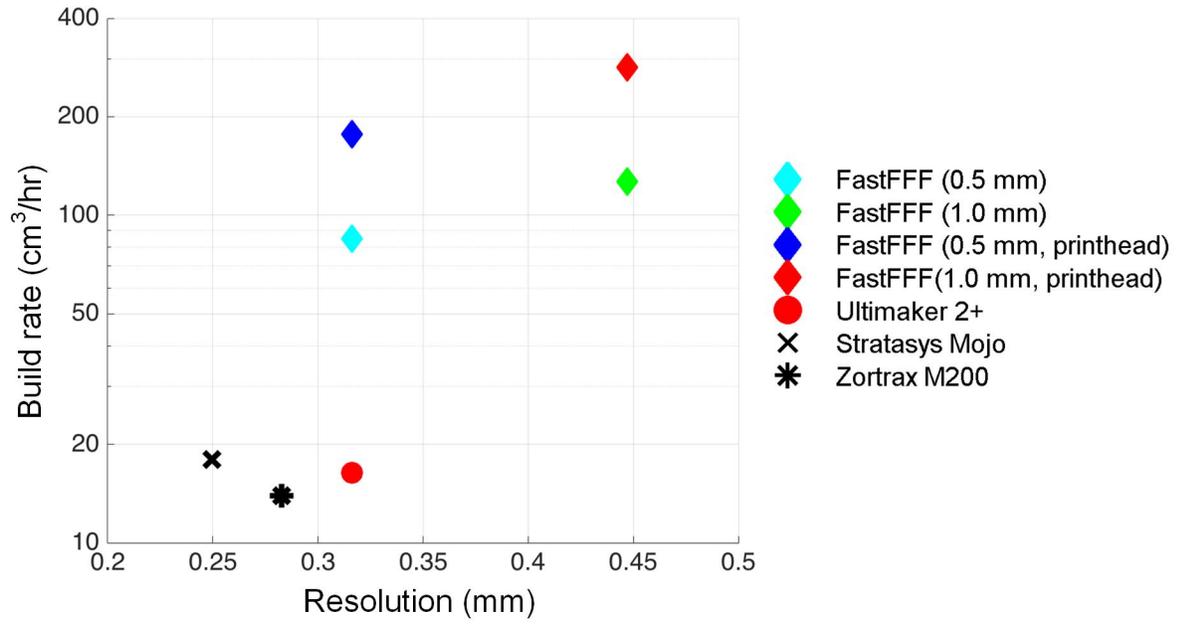

**Figure 5.** Benchmarking of FastFFF system performance, in terms of volumetric build rate and resolution, versus commercial desktop FFF systems.